\documentclass[]{article}
\usepackage{spconf,amsmath,graphicx}

\usepackage{latexsym}
\usepackage{amsmath,amssymb}
\usepackage{enumerate}
\usepackage{comment}
\usepackage{color}
\usepackage{bm}
\usepackage{array}
\usepackage{multirow}
\usepackage{colortbl}
\usepackage{subfigure}
\usepackage{setspace}

\newlength\savedwidth
\newcommand{\wcline}[1]{\noalign{\global\savedwidth\arrayrulewidth\global\arrayrulewidth 1.0pt} \cline{#1}
\noalign{\global\arrayrulewidth\savedwidth}}

\title{Joint Analysis of Acoustic Scenes and Sound Events Based on\\Multitask Learning with Dynamic Weight Adaptation}

\name{Kayo Nada, Keisuke Imoto, and Takao Tsuchiya}
\address{Faculty of Science and Engineering, Doshisha University, Japan.}

\begin{document}
\maketitle
%
\begin{abstract}
Acoustic scene classification (ASC) and sound event detection (SED) are major topics in environmental sound analysis. Considering that acoustic scenes and sound events are closely related to each other, the joint analysis of acoustic scenes and sound events using multitask learning (MTL)-based neural networks was proposed in some previous works. Conventional methods train MTL-based models using a linear combination of ASC and SED loss functions with constant weights. However, the performance of conventional MTL-based methods depends strongly on the weights of the ASC and SED losses, and it is difficult to determine the appropriate balance between the constant weights of the losses of MTL of ASC and SED. In this paper, we thus propose dynamic weight adaptation methods for MTL of ASC and SED based on dynamic weight average and multi--focal loss to adjust the learning weights automatically. Evaluation experiments using parts of the TUT Acoustic Scenes 2016/2017 and TUT Sound Events 2016/2017 are conducted, and we show that the proposed methods improve the scene classification and event detection performance characteristics compared with the conventional MTL-based method. We then investigate how the learning weights of ASC and SED tasks dynamically adapt as the model training progresses.
\end{abstract}
%
%
%
%
\section{Introduction}
\label{sec:intro}
The automatic analysis of environmental sounds has recently received increasing interest from the research and industrial communities.
Applications of environmental sound analysis cover a wide range from machine condition monitoring systems \cite{Chakrabarty_ICASSP2016_01,Koizumi_DCASE2020_01}, to automatic surveillance \cite{Chan_EUSIPCO2010_01}, automatic life-logging \cite{Stork_ROMAN2012_01,Imoto_INTERSPEECH2013_01}, media retrieval \cite{Fonseca_DCASE2018_01}, and biomonitoring systems \cite{Salamon_PLoSOne2016_01,Morfi_JASA2021_01,Morfi_DCASE2021_01}.

Environmental sound sound analysis has several subtasks, and acoustic scene classification (ASC) and sound event detection (SED) are often addressed as the primary tasks.
ASC is the subtask that estimates a predefined acoustic scene class, such as ``office,'' ``residential area,'' ``train,'' or ``indoor,'' from an acoustic signal. 
The SED subtask detects sound event labels and their start and end times from an audio recording, where a sound event represents the type of sound, such as ``keyboard typing,'' ``car,'' ``cutlery,'' or ``people talking.''

In recent years, many neural-network-based methods such as convolutional neural networks (CNNs), convolutional recurrent neural networks (CRNNs), and Transformers have been applied in ASC and SED \cite{Valenti_IJCNN2017_01,Liping_DCASE2018_01,Tanabe_DCASE2018_01,Raveh_DCASE2018_01,Hershey_ICASSP2017_01,Cakir_TASLP2017_01,Kong_TASLP2020_01,Miyazaki_DCASE2020_01}. 
For instance, Valenti et al. proposed a scene classification method using a simple CNN architecture \cite{Valenti_IJCNN2017_01}.
More recently, Liping et al. \cite{Liping_DCASE2018_01}, Tanabe et al. \cite{Tanabe_DCASE2018_01}, and Raveh and Amar \cite{Raveh_DCASE2018_01} have respectively proposed ASC methods based on Xception, VGG, and ResNet, which are widely used in image recognition.
Hershey et al. proposed a CNN-based event detection method \cite{Hershey_ICASSP2017_01}, and \c{C}ak\i r et al. applied a CRNN, which can capture temporal information on sound events, to SED \cite{Cakir_TASLP2017_01}. 
Furthermore, Kong et al. \cite{Kong_TASLP2020_01} and Miyazaki et al. \cite{Miyazaki_DCASE2020_01} proposed Transformer-based SED and Conformer-based SED, respectively.

In these conventional methods of environmental sound analysis, acoustic scenes and sound events are analyzed separately; however, acoustic scenes and sound events are related to each other.
For example, in the acoustic scene of ``home,'' the sound events ``dishes'' and ``glass jingling'' are more likely to occur, while the sound events such as ``cars'' and ``birds singing'' are less likely to occur. 
Therefore, when recognizing the sound events of ``dishes'' and ``glass jingling,'' information on the acoustic scene ``home'' helps identify these sound events, and vice versa.

On the basis of this observation, Mesaros et al. \cite{Mesaros_EUSIPCO2011_01} and Heittola et al. \cite{Heittola_JASM2013_01} have proposed methods of SED that take into account acoustic scene information in an unsupervised manner.
Imoto and Shimauchi \cite{Imoto_IEICE2016_01} and Imoto and Ono \cite{Imoto_TASLP2019_01} have proposed scene classification methods that consider sound event information using Bayesian generative models.
Bear et al. \cite{Bear_INTERSPEECH2019_01}, Tonami et al. \cite{Tonami_IEICE2021_01}, and Imoto et al. \cite{Imoto_ICASSP2020_01} have proposed the joint analysis methods for acoustic scenes and sound events utilizing MTL-based neural network models of ASC and SED.

These methods train MTL model parameters by linearly combining ASC and SED losses with constant weights. 
However, the studies using conventional methods showed that scene classification and event detection performance characteristics depend on the constant weights of ASC and SED losses, and that the determination of appropriate ASC and SED loss weights is difficult.
In addition, it may be desirable to dynamically change the learning weights as model training progresses.
Therefore, we have proposed a dynamic weight adaptation method for the multitask learning of ASC and SED based on multi--focal loss (MFL) \cite{Nada_APSIPA2021_01}.
In this paper, we newly introduce another dynamic weight adaptation technique using the dynamic weight average (DWA) \cite{Liu_CVPR2019_01} to MTL of ASC and SED in addition to a more detailed discussion of the dynamic weight adaptation method based on MFL.
Moreover, we evaluate ASC and SED performance characteristics in detail and disclose how the weights of MTL of ASC and SED affect their performance characteristics.

\begin{table*}[t!]
\small
\centering
\caption{Co-occurrence of acoustic scenes and sound events in training dataset used in evaluation experiments (TUT Acoustic Scenes 2016 and TUT Sound Events 2016/2017 [29], [30])}
\label{tbl:event_occurrence}
\vspace{5pt}
\begin{tabular}{c}
\includegraphics[width=2.02\columnwidth]{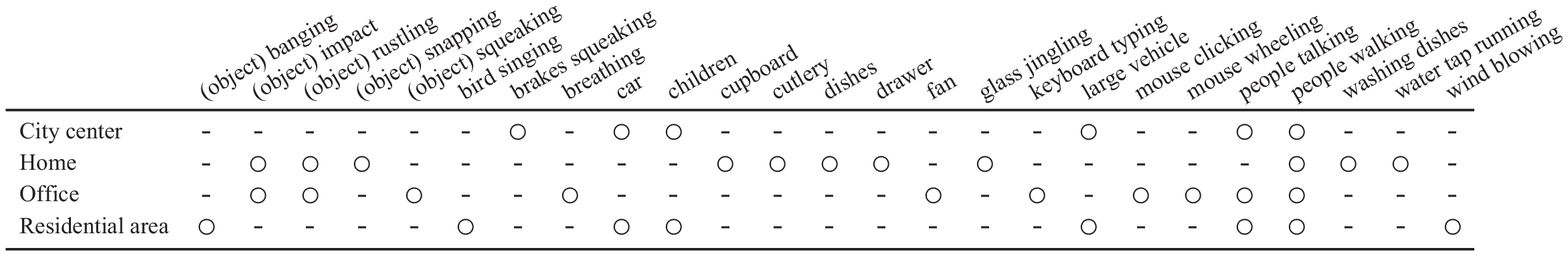}
\end{tabular}
\vspace{5pt}
\end{table*}

The rest of this paper is organized as follows.
In section 2, we discuss the conventional methods of ASC, SED, and joint analysis of ASC and SED using a MTL-based neural network.
In section 3, we propose dynamic adaptive methods for loss weights in multitask learning based on DWA and MFL.
In section 4, we describe our experiments conducted to evaluate the performance of the proposed method.
Finally, we conclude this work in section 5.
%
%
\section{Conventional Methods}
\label{sec:conventional}
\subsection{ASC and SED Methods Based on Neural Networks}
\label{ssec:ConvASCandSED}
In this section, we overview the conventional scene classification and event detection methods based on neural networks. 
Many conventional methods of ASC and SED apply a neural network framework, such as a convolutional neural network (CNN) \cite{Valenti_IJCNN2017_01,Hershey_ICASSP2017_01}, a recurrent neural network (RNN) \cite{Cakir_TASLP2017_01}, or a Transformer \cite{Kong_TASLP2020_01}. 

Suppose that an acoustic feature $X \in \mathcal{R}^{D \times L}$, such as the time series of mel frequency cepstrum coefficients (MFCCs) or the log mel-band spectrogram, is calculated from an observed acoustic signal.
Here, we suppose that $D$ and $L$ are the dimension of the acoustic feature and the number of time frames of the input acoustic feature, respectively.
The conventional methods of ASC and SED feed the acoustic feature $X$ into the network.
The ASC and SED networks have the convolution and/or recurrent layers, as well as the Transformer layer, which is followed by a softmax layer for ASC or a sigmoid layer for SED.

In ASC, the model parameters are tuned using the network output and the following cross-entropy (CE) loss function ${\mathcal L}_{{\rm scene}}$:

\begin{align}
{\mathcal L}_{{\rm scene}} = - \sum^{N}_{n=1} {\Big \{} z_{n} \log ( y_{n} ) {\Big \}},
\label{eq:scene_loss}
\end{align}

\noindent where $N$, $z_{n}$, and $y_{n}$ are the number of acoustic scene classes, the target scene label of acoustic scene $n$, and the network output, respectively.
If the input acoustic signal is most associated with acoustic scene $n$, the target scene label $z_{n}$ equals to 1 and 0 otherwise.

On the other hand, in SED, sound events may overlap in the time axis.
Thus, the model parameters are tuned utilizing the network output and the following binary cross-entropy (BCE) loss function ${\mathcal L}_{{\rm event}}$:

\begin{align}
&{\mathcal L}_{{\rm event}} \nonumber\\
&\hspace{5pt} = - \sum^{L}_{l=1} {\Big \{} {\bf z}_{l} \log ( {\bf y}_{l} ) + (1 - {\bf z}_{l}) \log (1 - {\bf y}_{l} ) {\Big \}} \nonumber\\[3pt]
&\hspace{5pt} = - \! \sum^{L, \hspace{0.5pt} M}_{l,m=1} \hspace{-2pt} {\Big \{} z_{l,m} \log ( y_{l,m} ) + (1 - z_{l,m}) \log (1 - y_{l,m}) {\Big \}},\nonumber\\[-2pt]
\label{eq:event_loss}
\end{align}

\noindent where $T$, $M$, $z_{l,m}$, and $y_{l,m}$ are the number of time frames in the acoustic feature $X$, the number of sound event classes, the target event label $m$ in time frame $l$, and the network output for sound event $m$ in time frame $l$, respectively.
\begin{figure}[t!]
\centering
\includegraphics[width=0.94\columnwidth]{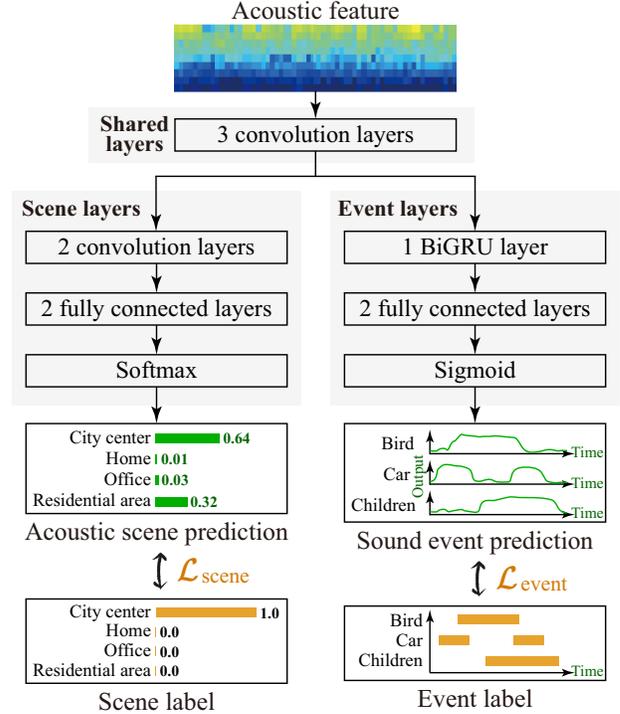}
\caption{Network structure of conventional MTL-based method of ASC and SED \cite{Tonami_IEICE2021_01,Imoto_ICASSP2020_01}}
\label{fig:conventionalMTL}
\vspace{6pt}
\end{figure}
%
%
%
\subsection{Conventional Joint Analysis Method for Acoustic Scenes and Sound Events Based on MTL}
\label{ssec:ConvMTL}
In many conventional methods, acoustic scenes and sound events are analyzed separately.
However, acoustic scenes and sound events are closely related and tend to co-occur.
Table~\ref{tbl:event_occurrence} shows that the occurrence of sound events is biased toward acoustic scenes; thus, sound events and acoustic scenes will be beneficial in their mutual estimation. 
On the basis of this idea, joint analysis methods for acoustic scenes and sound events using MTL networks have been proposed \cite{Bear_INTERSPEECH2019_01,Tonami_IEICE2021_01,Imoto_ICASSP2020_01}.

In MTL-based methods, a network is partially shared to hold common information on acoustic scenes and sound events in the shared layers, as shown in Fig.~\ref{fig:conventionalMTL}.
The output of the shared layer is then fed to the layers dedicated to ASC and SED, which consists of, for example, the CNN and bidirectional gated recurrent unit (BiGRU) layers for scene classification and event detection, respectively.
To train the model parameters, the conventional method \cite{Tonami_IEICE2021_01,Imoto_ICASSP2020_01} utilizes the following loss function that linearly combines ASC and SED loss functions with constant weights $\lambda_{1}$ and $\lambda_{2}$:

\vspace{-3pt}
\begin{align}
{\mathcal L}_{{\rm MTL}} &= \lambda_{1} {\mathcal L}_{{\rm scene}} + \lambda_{2} {\mathcal L}_{{\rm event}}.
\label{eq:mtl_loss}
\end{align}
\vspace{3pt}
%

%
%
\section{Proposed Method}
\label{sec:proposed}
\subsection{Motivation}
\label{ssec:motivation}
The evaluation experiments using conventional MTL-based methods disclosed that the MTL framework indeed improves the ASC and SED performance characteristics compared with the case of the ASC and SED methods based on single-task networks \cite{Tonami_IEICE2021_01}.
However, the performance characteristics of ASC and SED depend on the constant weights $\lambda_{1}$ and $\lambda_{2}$, and determining appropriate weights of ASC and SED losses is not easy.
Furthermore, in conventional MTL-based methods, $\lambda_{1}$ and $\lambda_{2}$ are constant during model training. 
However, it may be desirable to dynamically change the balance of weights according to the progress of model training of ASC and SED.
To overcome this limitation of conventional MTL-based methods, we apply dynamic weight adaptation methods for MTL of ASC and SED based on the dynamic weight average \cite{Liu_CVPR2019_01} and MFL \cite{Lin_ICCV2017_01}.
%
%
%
%
\subsection{Dynamic Weight Adaptation Based on Dynamic Weight Average}
\label{ssec:dwa}
We first apply DWA \cite{Liu_CVPR2019_01} to automatically determine the appropriate weights of ASC and SED losses in MTL.
DWA averages the training weights over the model training process by calculating the rate of change in training loss for each task as follows:

\begin{align}
\lambda_{k} (t) &= \frac{K \exp (w_{k}(t-1)/T)}{\sum_{i} \exp (w_{i}(t-1)/T)}\\[2pt]
w_{k} (t-1) &= \frac{\mathcal{L}_{k} (t-1)}{\mathcal{L}_{k} (t-2)},
\label{eq:dwa}
\end{align}
\vspace{1pt}

\noindent where $K$, $k$, $t$, and $T$ are the number of tasks in MTL, the task index, the iteration index of model training, and the temperature parameter that controls the change rate of loss weights, respectively.
$w_{k}()$ indicates the change rate of the training loss in the $k$th task.

In our work, we consider that the analysis of each acoustic scene and sound event is a separated task and uses the following loss function for model training:

\vspace{0pt}
\begin{align}
{\mathcal L}_{{\rm MTL}} &= {\mathcal L}_{{\rm scene}} + {\mathcal L}_{{\rm event}}\nonumber\\
&= - \sum^{N}_{n=1} \lambda_{n} {\Big \{} z_{n} \log ( y_{n} ) {\Big \}} \nonumber\\[6pt]
&\hspace{12pt} - \!\! \sum^{L, \hspace{0.5pt} M}_{l,m=1} \!\! \lambda_{m} {\Big \{} z_{l,m} \log ( y_{l,m} ) \nonumber\\
&\hspace{25pt} + (1 - z_{l,m}) \log (1 - y_{l,m}) {\Big \}},
\end{align}
\vspace{5pt}

\noindent where $\lambda_{n}$ and $\lambda_{m}$ are the training weights for acoustic scene $n$ and sound event $m$, respectively.
%
%
%
\subsection{Dynamic Weight Adaptation of MTL of ASC and SED Based on Multi--focal Loss}
\label{ssec:mfl}
We can alternatively apply a focal loss \cite{Lin_ICCV2017_01,Noh_Sensors2020_01} to dynamically adapt the weights of ASC and SED losses in MTL.
The focal loss was originally developed in order to dynamically adjust the training weight according to the difficulty/ease of model training as follows:

\vspace{-3pt}
\begin{align}
{\mathcal L} &= - \! \sum^{N}_{n=1} {\Big \{} (1 - y_{n} )^{\hspace{-0.3pt} \eta} \hspace{0.5pt} z_{n} \log ( y_{n} ) {\Big \}},
\label{eq:scene_focal_loss_01}
\end{align}
\vspace{3pt}

\noindent where $\eta$ is the constant focusing parameter.
That is, the focal loss automatically modifies the training weight according to the prediction error of each task.

In this work, we apply the focal loss to dynamically adjust the weight of MTL losses of ASC and SED, replacing ${\mathcal L}_{{\rm scene}}$ and ${\mathcal L}_{{\rm event}}$ in Eq.~(\ref{eq:mtl_loss}) with the following loss functions:

\begin{align}
\label{eq:scene_focal_loss_02}
{\mathcal L}_{{\rm scene}} &= - \sum^{N}_{n=1} {\Big \{} (1 - y_{n} )^{\hspace{-0.3pt} \eta} \hspace{0.5pt} z_{n} \log ( y_{n} ) {\Big \}},\\[4pt]
{\mathcal L}_{{\rm event}} &= - \! \sum^{L, \hspace{0.5pt} M}_{l,m=1} {\Big \{} (1 - y_{l,m} )^{\hspace{-0.3pt} \gamma} \hspace{0.5pt} z_{l,m} \log ( y_{l,m} )\nonumber\\
&\hspace{21pt} + y_{l,m}^{\zeta} (1 - z_{l,m}) \log (1 - y_{l,m}) {\Big \}},
\label{eq:event_focal_loss}
\end{align}
\vspace{1pt}

\noindent where $\gamma$ and $\zeta$ are the constant focusing parameters. 
The proposed MFL function automatically determines the appropriate balance of training weights between any classes of acoustic scenes and sound events.
%
%
%
\section{Evaluation Experiments}
\label{sec:experiment}
\subsection{Experimental Conditions}
\label{ssec:condition}
We conducted experiments to evaluate the performance of dynamic weight adaptation for MTL of ASC and SED based on DWA and MFL.
For the evaluation, we used a dataset that consists of parts of the TUT Acoustic Scenes 2016/2017 and TUT Sound Events 2016/2017 \cite{Mesaros_EUSIPCO2016_01,Mesaros_DCASE2017_01}.
From the TUT datasets, we chose sound clips including four acoustic scenes, ``city center,'' ``home,'' ``office,'' and ``residential area,'' for a total of 266 min of sounds (192 min for the development set and 74 min for the evaluation set). 
The datasets include the 25 different sound events listed in Table~\ref{tbl:event_occurrence}.
Details of the datasets used for the evaluation experiments are found in \cite{Imoto_dataset2019_01}. 

As an acoustic feature, 64-dimensional log mel-band energy, which was extracted every 40 ms with a hop size of 20 ms, was applied.
We then fed the acoustic feature to the MTL network applied in \cite{Tonami_IEICE2021_01}.

The parameters for MTL loss in MFL are set to 1) $\lambda_{1} = \lambda_{2} = \eta = \gamma = \zeta = 1.0$ (MTL w/ MFL 1) and 2) $\lambda_{1}=0.001$, $\lambda_{2}=1.0$, $\eta=1.0$, $\gamma=1.0$, and $\zeta=0.0625$ (MTL w/ MFL 2), which is selected by referring to \cite{Imoto_ICASSP2021_01}.
For each method, we evaluated the ASC and SED performance characteristics with 20 random initial values of the model parameters.
The details of the network structure and other experimental conditions are listed in Tables~\ref{tbl:networks} and \ref{tbl:parameter}.
\begin{table}[t]
\footnotesize
\caption{Details of network structure of MTL of ASC and SED}
\label{tbl:networks}
\begin{center}
\begin{tabular}{ccc}
\wcline{1-3}
\!\!&\!\!\\[-7.5pt]
\multicolumn{3}{c}{\textbf{Shared network}}\\
\wcline{1-3}
\!\!&\!\!\\[-7.5pt]
\multicolumn{3}{c}{Log-mel energy}\\[0pt]
\multicolumn{3}{c}{500 frames $\times$ 64 mel bin}\\[0pt]
\cline{1-3}
\!&\\[-7.5pt]
\multicolumn{3}{c}{3$\times$3 kernel size/128 ch.}\\[0pt]
\multicolumn{3}{c}{Batch norm., Leaky ReLU}\\[0pt]
\multicolumn{3}{c}{1$\times$8 Max pooling}\\[0pt]
\cline{1-3}
\!&\\[-7.5pt]
\multicolumn{3}{c}{$\begin{pmatrix} \textrm{3$\times$3 kernel size/128 ch.}\\
\textrm{Batch norm., Leaky ReLU}\\
\textrm{1$\times$2 Max pooling}
\end{pmatrix}$ $\times$ 2
}\\[0pt]
\!&\\[-7.5pt]
\wcline{1-3}
\multicolumn{1}{c}{}\\[-7.5pt]
\multicolumn{1}{c}{\textbf{Scene layers}}&\ \ \ &\textbf{Event layers}\\
\wcline{1-1}\wcline{3-3}
\multicolumn{1}{c}{}\\[-7.5pt]
\multicolumn{1}{c}{3$\times$3 kernel size/256 ch.}&\ \ \ &\\
\multicolumn{1}{c}{Batch norm., Leaky ReLU}&\ \ \ &BiGRU w/ 32 units\\
\multicolumn{1}{c}{25$\times$1 Max pooling}&\ \ \ &\\
\cline{1-1} \cline{3-3}
\multicolumn{1}{c}{}\\[-7.5pt]
\multicolumn{1}{c}{3$\times$3 kernel size/256 ch.}&\ \ \ &\multirow{2}{*}{}\\
\multicolumn{1}{c}{Batch norm., Leaky ReLU}&&FC w/ 32 units, Leaky ReLU\\
\multicolumn{1}{c}{Global max pooling}&\ \ \ &\\
\cline{1-1} \cline{3-3}
\multicolumn{1}{c}{}\\[-7.5pt]
\multicolumn{1}{c}{FC w/ 32 units, Leaky ReLU}&\ \ \ &FC w/ 25 units, sigmoid\\
\cline{1-1}\wcline{3-3}
\multicolumn{1}{c}{}\\[-8pt]
\multicolumn{1}{c}{FC w/ 4 units, Softmax}&&\\
\wcline{1-1}
\end{tabular}
\end{center}
\end{table}
\begin{table}[t]
\small
\centering
\caption{Experimental conditions}
\label{tbl:parameter}
\begin{tabular}{ll}
\wcline{1-2}
&\\[-7.2pt]
Acoustic feature \hspace{7pt}&\hspace{7pt} Log-mel energy \\
Dim. of acoustic feature \hspace{7pt}&\hspace{7pt} 64\\
Frame length/shift \hspace{7pt}&\hspace{7pt} 40 ms/20 ms\\
Length of sound clip \hspace{7pt}&\hspace{7pt} 10 s\\
Optimizer \hspace{7pt}&\hspace{7pt} RAdam \cite{Liu_ICLR2020_01}\\[0pt]
Temperature parameter $T$ \hspace{7pt}&\hspace{7pt} 1.0\\[0pt]
\wcline{1-2}
\end{tabular}
\vspace{5pt}
\end{table}
\begin{table*}[t]
\small
\centering
\caption{Performances of scene classification and event detection}
\label{tbl:performance01}
\begin{tabular}{lccccccc}
\wcline{1-6}
&\\[-7.2pt]
&\multicolumn{2}{c}{\textbf{Scene classification}}&&\multicolumn{2}{c}{\textbf{Event detection}} \\
\cline{2-3}\cline{5-6}
&\\[-8.2pt]
\multicolumn{1}{c}{\multirow{-2.1}{*}{\textbf{Method}}} & Micro-Fscore \hspace{5pt}&\hspace{5pt} Macro-Fscore \hspace{5pt}&\ \ \ &\hspace{5pt} Micro-Fscore \hspace{5pt}&\hspace{5pt} Macro-Fscore\\
\wcline{1-6}
&\\[-7.2pt]
CNN (ASC) & 85.00\% & 84.29\% && - & - \\
CNN-BiGRU (SED) & - & - && 42.54\% & 11.20\% \\

Conventional MTL {\footnotesize ($\lambda_{1}\!=\!0.001, \lambda_{2}\!=\!1.0$)} \!& 87.64\% & 87.50\% && 42.78\% & 11.24\%\\
MTL w/ DWA {\footnotesize ($T = 1.0$)} & \textbf{90.87\%} & \textbf{90.62\%} && 45.48\% & 11.71\%\\
MTL w/ MFL 1 {\footnotesize ($\lambda_{1} = \lambda_{2} = \gamma = \zeta = \eta = 1.0$)}& 90.51\% & 90.26\% && 44.40\% & 12.04\%\\
MTL w/ MFL 2 {\footnotesize ($\lambda_{1}=0.001$, $\lambda_{2}=\eta=\gamma=1.0$, $\zeta=0.0625$)} & 89.56\% & 89.35\% && \textbf{46.06\%} & \textbf{12.73\%}\\
\wcline{1-6}
\end{tabular}
\end{table*}
\begin{table*}[t]
\footnotesize
\centering
\caption{Average Fscores for selected sound events}
\label{tab:performance02}
\begin{tabular}{lcccccccccc}
\wcline{1-11}\\
&\\[-16pt]
& {\bf  Bird} \ &\ {\bf Brakes} \ &\ \ &\ \ &\ \ &\ \ &\ {\bf Keyboard} \ &\ {\bf People} \ &\ {\bf Washing} \ &\ {\bf Water tap}\ \\
\multicolumn{1}{c}{\multirow{-1.9}{*}{\bf Method}} \ &\ {\bf singing} \ &\ {\bf squeaking} \ &\ \multirow{-1.9}{*}{\bf  Car} \ &\ \multirow{-1.9}{*}{\bf Cupboard} \ &\ \multirow{-1.9}{*}{\bf Dishes} \ &\ \multirow{-1.9}{*}{\bf Fan} \ &\ {\bf typing}\ &\ {\bf walking} \ &\ {\bf dishes} \ &\ {\bf running}\ \\
&\\[-8pt]
\wcline{1-11}\\[-7pt]
\ CNN-BiGRU & \textbf{46.36\%} & 0.00\% & 44.53\% & 0.00\% & 0.17\% & 88.11\% & \ \ 4.34\% & \textbf{12.55\%} & 21.29\% & 25.42\%\ \\
\ Conv. MTL & 46.29\% & 0.00\% & \textbf{45.51\%} & 0.00\% & 0.25\% & 87.68\% & \ \ 5.08\% & 10.46\% & 21.03\% & 29.20\%\ \\
\ MTL w/ DWA & 33.72\% & 0.18\% & 44.02\% & 0.00\% & 2.44\% & 91.09\% & \ \ 9.31\% & \ \ 8.69\% & 15.70\% & 63.80\%\ \\
\ MTL w/ MFL 1 & 33.81\% & 0.22\% & 43.08\% & 0.00\% & 1.47\% & 91.00\% & 10.06\% & 8.16\% & \textbf{24.94\%} & 64.83\%\ \\
\ MTL w/ MFL 2 & 35.48\% & \textbf{0.95\%} & 45.36\% & 0.00\% & \textbf{3.97\%} & \textbf{93.56\%} & \textbf{11.48\%} & 10.01\% & 18.36\% & \textbf{66.72\%}\ \\
\wcline{1-11}
\end{tabular}
\vspace{10pt}
\end{table*}
%
%
%
%
\subsection{Experimental Results}
\label{ssec:result}
\subsubsection{Overall Performance of ASC and SED}
\label{sssec:result1}
Table~\ref{tbl:performance01} shows the average performance characteristics of scene classification and event detection by the conventional and proposed methods.
For CNN (ASC) and CNN-BiGRU (SED), we applied the same network structures with shared + scene layers and shared + event layers, respectively.
The results show that the proposed dynamic weight adaptation methods using DWA and MFL achieve higher performances in both scene classification and event detection than the conventional methods.
In particular, when applying MTL w/ MFL 2 ($\lambda_{1}=0.001$, $\lambda_{2}=1.0$, $\eta=1.0$, $\gamma=1.0$, and $\zeta=0.0625$), micro-Fscores for scene classification and event detection are improved by 1.92 and 3.28 percentage points compared with those of the conventional MTL-based method.
%
%
\subsubsection{Detailed Results of Scene Classification and Event Detection}
\label{sssec:detailedresults}
\begin{figure}[t!]
\vspace{5pt}
\centering
  \hspace{-1.0pt}
  \subfigure[CNN (ASC)]{%
  \includegraphics[width=0.51\columnwidth]{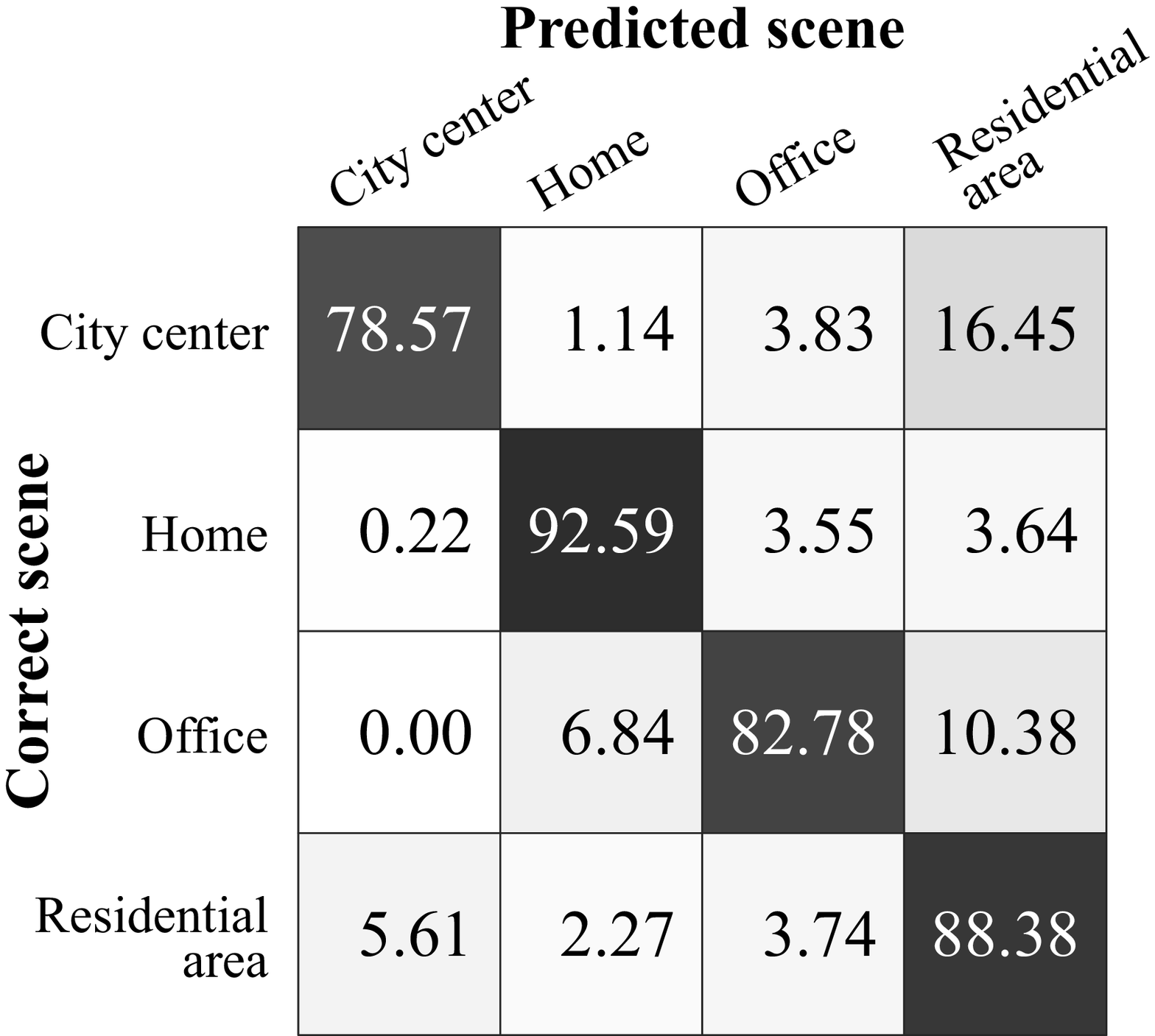}}%
  \subfigure[Conv. MTL]{%
  \includegraphics[width=0.51\columnwidth]{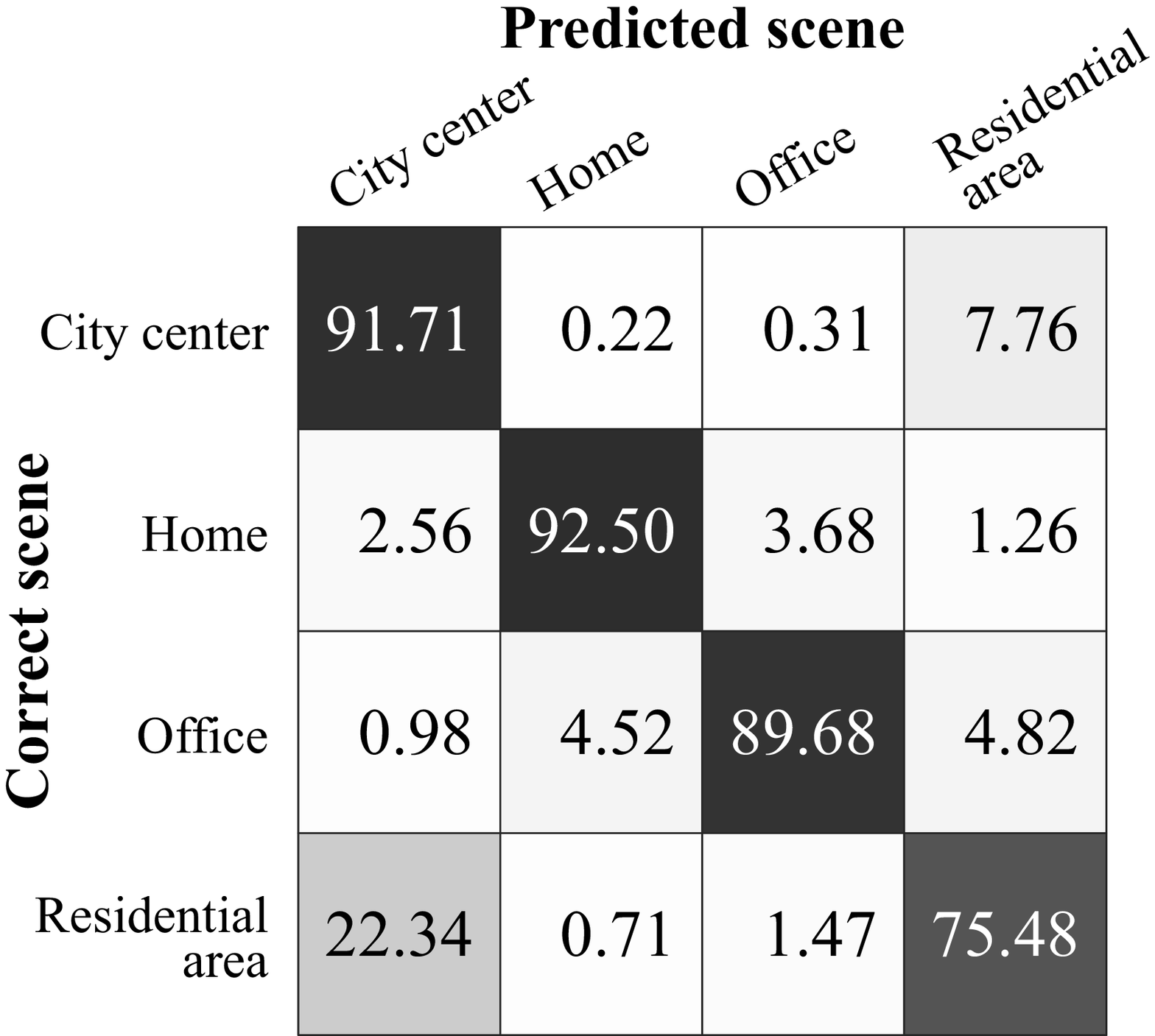}}%
  \ \\[3pt]
  \hspace{-1.0pt}
  \subfigure[MTL w/ DWA]{%
  \includegraphics[width=0.51\columnwidth]{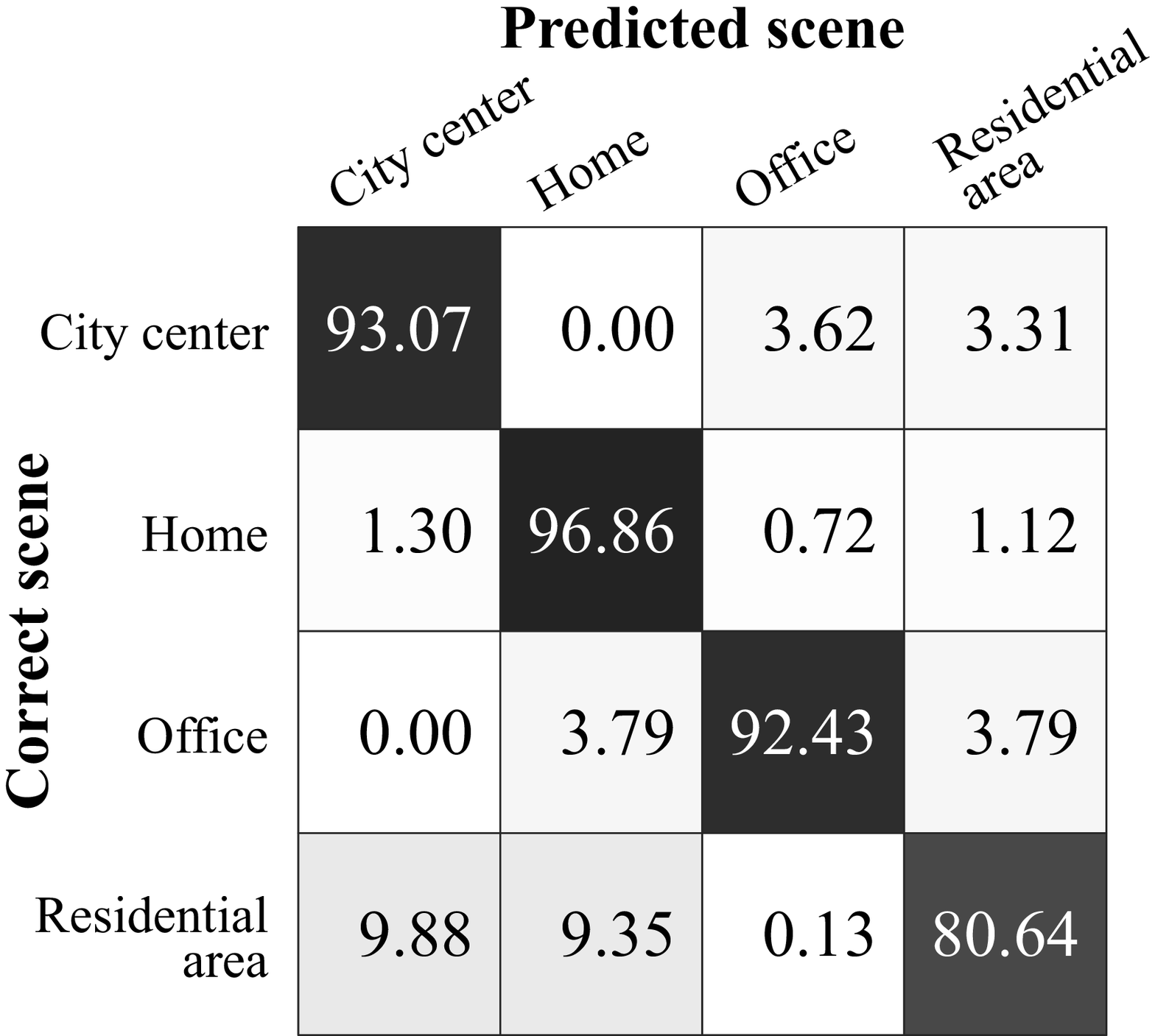}}%
  \subfigure[MTL w/ MFL 2]{%
  \includegraphics[width=0.51\columnwidth]{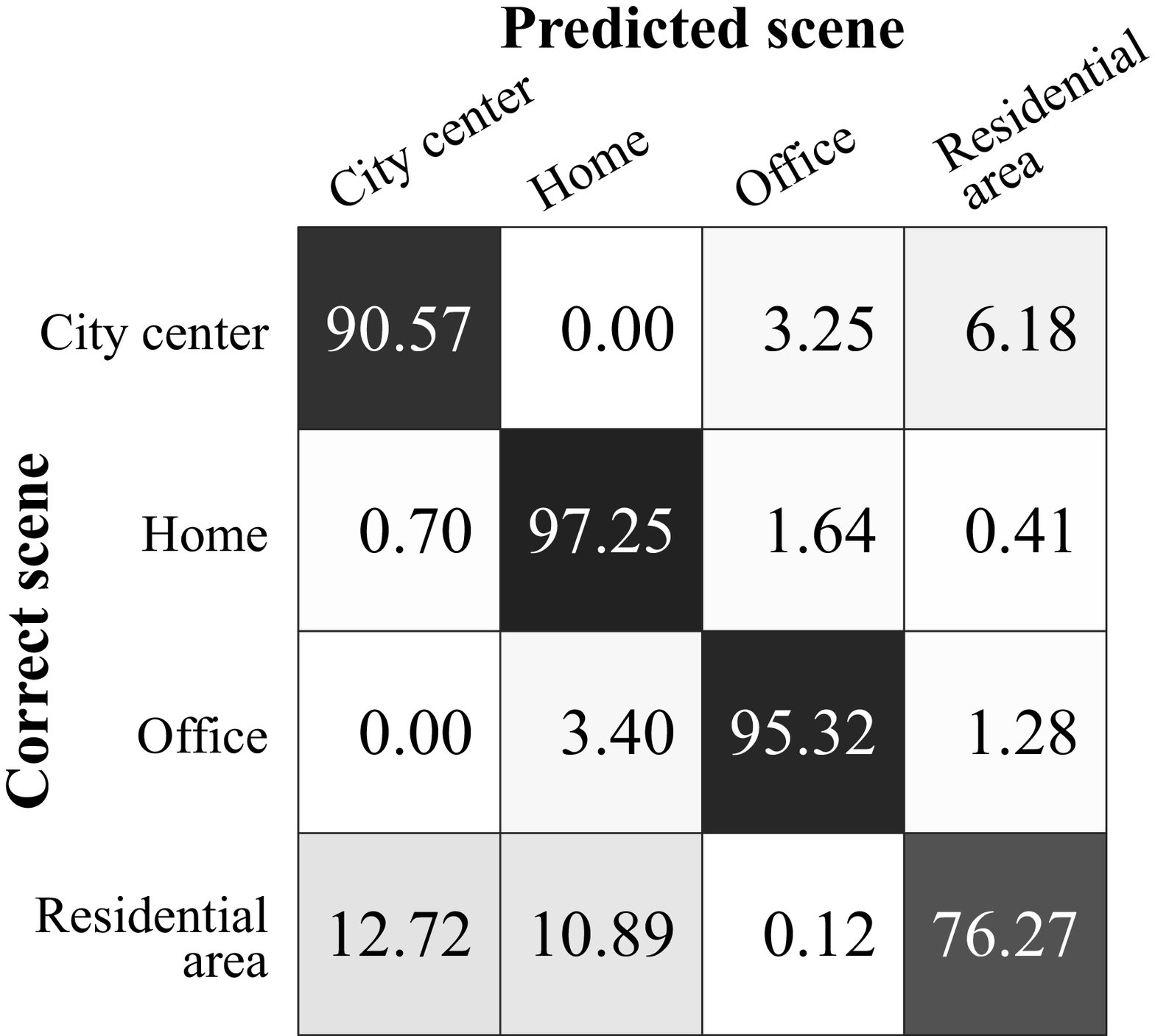}}%
\caption{Average scene classification results in terms of recall (\%)}
\label{fig:sceneres01}
\end{figure}

To investigate the scene classification and event detection performance characteristics in detail, we list the recalls and F-scores for each acoustic scene and sound event in Fig.~\ref{fig:sceneres01} and Table~\ref{tab:performance02}, respectively.
Figure~\ref{fig:sceneres01} shows that the performance improvement of the proposed MTL w/ DWA and MFL is balanced among many acoustic scenes.
This means that the model training of the ASC task was well-balanced, with less bias toward any particular scene.
\begin{figure}[t!]
\centering
\includegraphics[width=1.0\columnwidth]{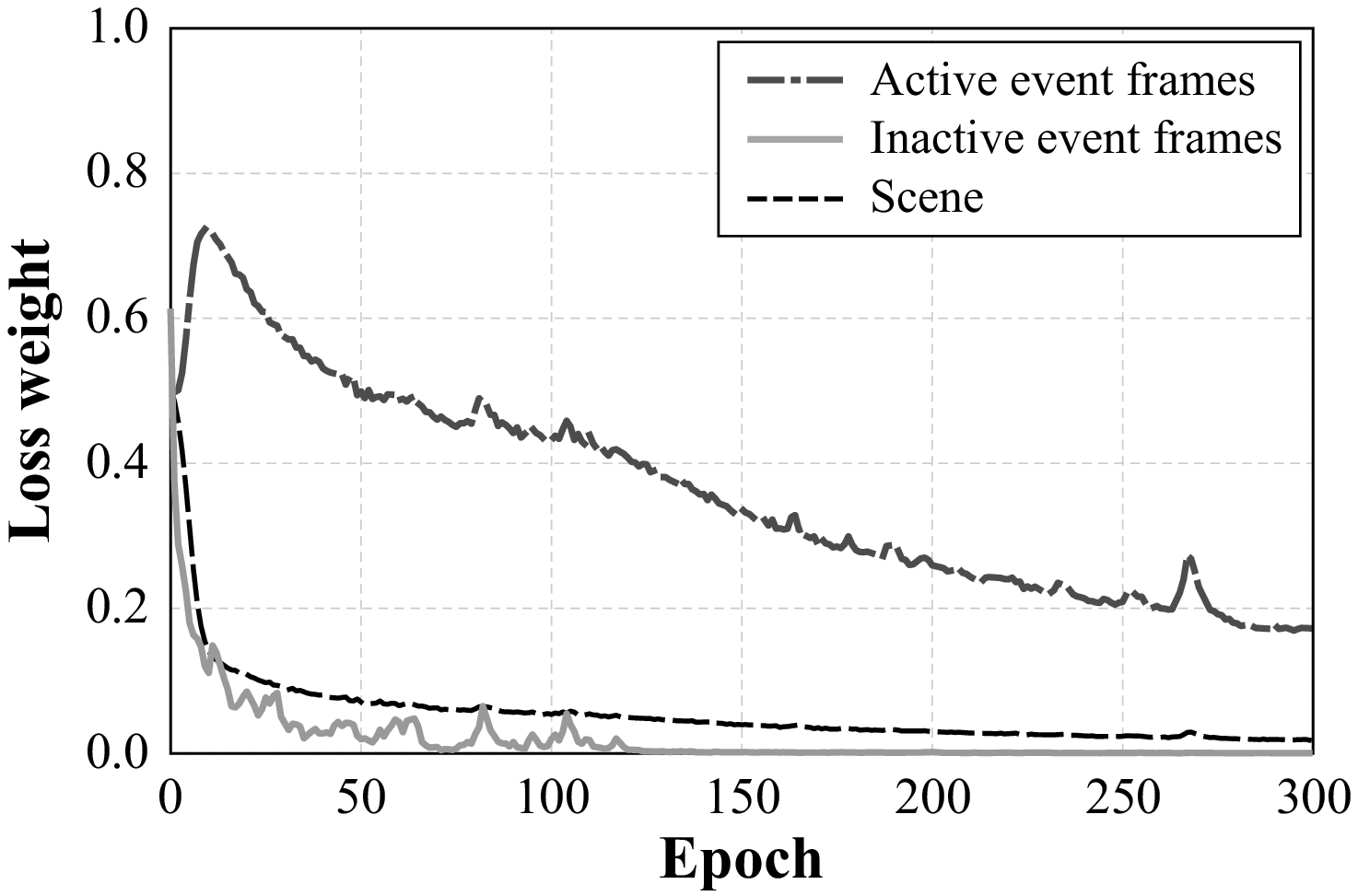}
\caption{Time variation of loss weights in MTL w/ MFL 1 ($\lambda_{1} = \lambda_{2} = \gamma = \zeta = \eta = 1.0$)}
\label{fig:eq_coeff_curve}
\end{figure}
\begin{figure}[t!]
\centering
\vspace{5pt}
\includegraphics[width=1.0\columnwidth]{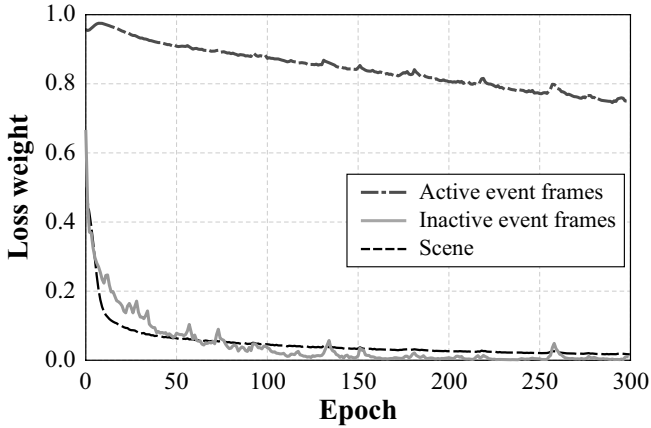}
\caption{Time variation of loss weights in MTL w/ MFL 2 ($\lambda_{1}=0.001$, $\lambda_{2}=\eta=\gamma=1.0$, $\zeta=0.0625$)}
\label{fig:conv_coeff_curve}
\end{figure}

Table~\ref{tab:performance02} shows that the proposed methods based on MTL w/ DWA and MFL improve Fscores for many sound events with lower detection performance when using the conventional MTL-based method, such as ``brakes squeaking,'' ``dishes,'' ``keyboard typing,'' and ``water tap running.''
On the other hand, the detection performance for sound events that can be detected accurately when using the conventional method (e.g., ``car'' and ``fan'') remains comparable.
This result indicates that the proposed method also enables the training of a well-balanced SED model in any sound events.
\begin{figure}[t!]
\centering
\includegraphics[width=0.99\columnwidth]{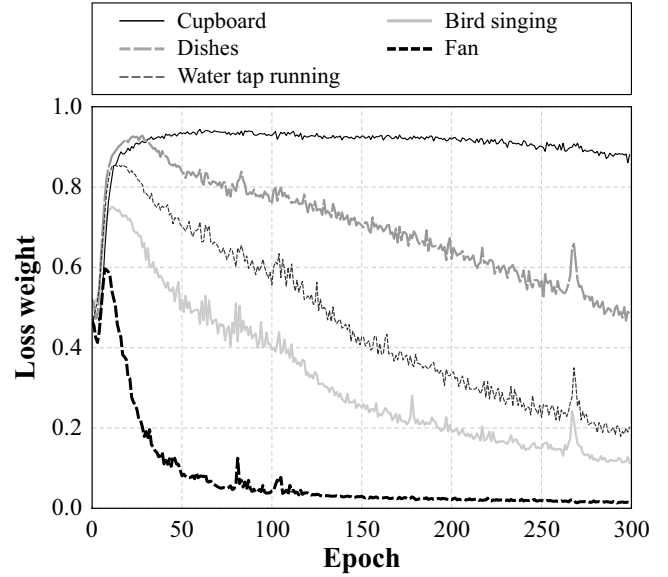}
\caption{Time variation of loss weights in MTL w/ MFL 1 for selected sound events ($\lambda_{1} = \lambda_{2} = \gamma = \zeta = \eta = 1.0$)}
\label{fig:eq_coeff_curve_event}
\vspace{10pt}
\end{figure}
\begin{figure}[t!]
\centering
\includegraphics[width=0.99\columnwidth]{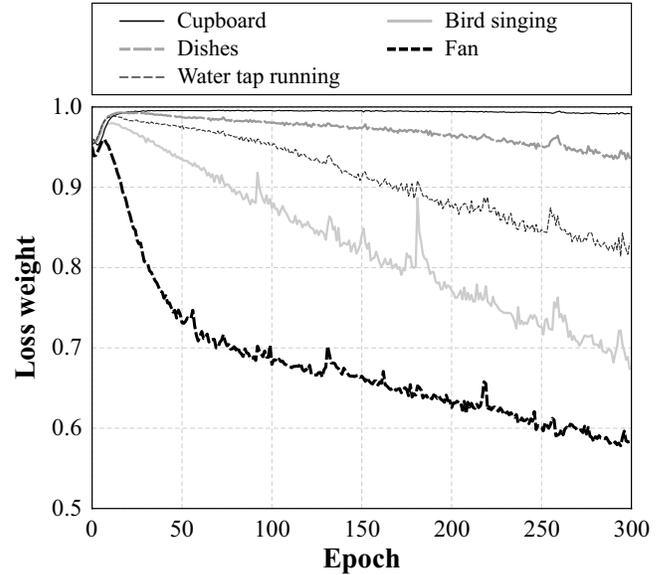}
\caption{Time variation of loss weights in MTL w/ MFL 2 for selected sound events ($\lambda_{1}=0.001$, $\lambda_{2}=\eta=\gamma=1.0$, $\zeta=0.0625$)}
\label{fig:conv_coeff_curve_event}
\end{figure}
%
%
%
\subsubsection{Time Variation of Loss Weights}
\label{sssec:time_variation}
We then investigate how the loss weights change in the model training with the proposed method.
In this experiment, we calculate the time variations of average loss weights in MTL w/ MFL for acoustic scene ($1 / N \cdot \sum^{N}_{n=1} (1 - y_{n} )^{\hspace{-0.3pt} \eta}$), active sound event frames ($1 / (LM) \cdot \sum^{L, \hspace{0.5pt} M}_{l,m=1} (1 - y_{l,m} )^{\hspace{-0.3pt} \gamma}$), and inactive sound event frames ($1 / (LM) \cdot \sum^{L, \hspace{0.5pt} M}_{l,m=1} y_{l,m}^{\zeta}$).
Figures~\ref{fig:eq_coeff_curve} and \ref{fig:conv_coeff_curve} show the time variations of loss weights in MTL w/ MFL 1 and MTL w/ MFL 2, respectively.
Both figures show similar trends, that is, the loss weight of acoustic scene decreases rapidly for a small number of epochs, whereas that of active sound event frames gradually decreases.
This implies that there is a difference in difficulty between ASC and SED, and ASC requires fewer epochs to train the model.
Thus, it is desirable to individually and dynamically change the loss weights as model training progresses, more specifically, setting the loss weights of ASC high only in the early stages of model training and reducing the weights significantly after that.
For inactive event frames, the loss weight becomes very small as the model training progresses.
This is because there are much larger inactive frames than active event frames, and the inactive frames tend to be stationary; thus, the model training with inactive event frames is easier than that with active event frames.
The proposed method can follow the behavior of the inactive frames and thus achieve higher performance than the conventional MTL-based method with the manually determined training weights.
For a detailed discussion on the effect of the active/inactive frames, refer to \cite{Imoto_ICASSP2021_01}.

To explore how the loss weight of each sound event changes during the model training, we also present the time variation of each sound event in MTL w/ MFL ($1 / L \cdot \sum^{L}_{l=1} (1 - y_{l})^{\hspace{-0.3pt} \gamma}$) as shown in Figs.~\ref{fig:eq_coeff_curve_event} and \ref{fig:conv_coeff_curve_event}.
These figures and Table~\ref{tab:performance02} indicate that there are differences in the difficulty of model training even among sound events, for example, for the sound event ``fan,'' the SED model is relatively easy to train but difficult to train for the sound event ``cupboard.''
On the other hand, the proposed method using MTL with MFL can adjust the learning rates automatically to some extent, successfully improving the SED performance for the sound event with which the model is difficult to train.
%
%
%
%
\section{Conclusions}
\label{sec:conclusion}
In this paper, we proposed the dynamic weight adaptation methods for MTL of ASC and SED.
In the proposed methods, we applied the MFL objective function and the DWA technique to ASC and SED losses to dynamically adapt the training weights of ASC and SED losses.
The proposed MFL and DWA can adapt the learning weights dynamically in accordance with the progress of model training.
We conducted the experiments using parts of the TUT Acoustic Scenes 2016/2017 and TUT Sound Events 2016/2017 datasets to evaluate the ASC and SED performance characteristics, which indicate that the proposed MTL of ASC and SED with MFL outperforms the conventional MTL method by 1.92 and 3.28 percentage points of the micro-Fscore in scene classification and event detection tasks, respectively.
Moreover, the experimental results also indicate that adapting the training weights dynamically in accordance with the progress of model training also contributes to improving the ASC and SED performances characteristics.
%
%
%
\section{Acknowledgement}
\label{sec:ack}
This work was supported by JSPS KAKENHI Grant Number JP20H00613.
%
%
%
\bibliographystyle{IEEEtran}
\bibliography{KeisukeImoto12,AST2022ref}

\begin{thebibliography}{10}
\providecommand{\url}[1]{#1}
\csname url@samestyle\endcsname
\providecommand{\newblock}{\relax}
\providecommand{\bibinfo}[2]{#2}
\providecommand{\BIBentrySTDinterwordspacing}{\spaceskip=0pt\relax}
\providecommand{\BIBentryALTinterwordstretchfactor}{4}
\providecommand{\BIBentryALTinterwordspacing}{\spaceskip=\fontdimen2\font plus
\BIBentryALTinterwordstretchfactor\fontdimen3\font minus
  \fontdimen4\font\relax}
\providecommand{\BIBforeignlanguage}[2]{{%
\expandafter\ifx\csname l@#1\endcsname\relax
\typeout{** WARNING: IEEEtran.bst: No hyphenation pattern has been}%
\typeout{** loaded for the language `#1'. Using the pattern for}%
\typeout{** the default language instead.}%
\else
\language=\csname l@#1\endcsname
\fi
#2}}
\providecommand{\BIBdecl}{\relax}
\BIBdecl

\bibitem{Chakrabarty_ICASSP2016_01}
D.~Chakrabarty and M.~Elhilali, ``Abnormal sound event detection using temporal
  trajectories mixtures,'' \emph{Proc. {IEEE} International Conference on
  Acoustics, Speech and Signal Processing {\rm (}ICASSP{\rm )}}, pp. 216--220,
  2016.

\bibitem{Koizumi_DCASE2020_01}
Y.~Koizumi, Y.~Kawaguchi, K.~Imoto, T.~Nakamura, Y.~Nikaido, R.~Tanabe,
  H.~Purohit, K.~Suefusa, T.~Endo, M.~Yasuda, and N.~Harada, ``Description and
  discussion on {DCASE2020} challenge task2: Unsupervised anomalous sound
  detection for machine condition monitoring,'' \emph{Proc. Detection and
  Classification of Acoustic Scenes and Events {\rm (}DCASE{\rm )}}, pp.
  81--85, 2020.

\bibitem{Chan_EUSIPCO2010_01}
C.~Chan and E.~W.~M. Yu, ``An abnormal sound detection and classification
  system for surveillance applications,'' \emph{Proc. European Signal
  Processing Conference {\rm (}EUSIPCO{\rm )}}, pp. 1851--1855, 2010.

\bibitem{Stork_ROMAN2012_01}
J.~A. Stork, L.~Spinello, J.~Silva, and K.~O. Arras, ``Audio-based human
  activity recognition using non-{M}arkovian ensemble voting,'' \emph{Proc.
  {IEEE} International Symposium on Robot and Human Interactive Communication
  {\rm (}RO-MAN{\rm )}}, pp. 509--514, 2012.

\bibitem{Imoto_INTERSPEECH2013_01}
K.~Imoto, S.~Shimauchi, H.~Uematsu, and H.~Ohmuro, ``User activity estimation
  method based on probabilistic generative model of acoustic event sequence
  with user activity and its subordinate categories,'' \emph{Proc.
  INTERSPEECH}, 2013.

\bibitem{Fonseca_DCASE2018_01}
E.~Fonseca, M.~Plakal, F.~Font, D.~P.~W. Ellis, X.~Favory, J.~Jordi, and
  X.~Serra, ``General-purpose tagging of freesound audio with {AudioSet}
  labels: Task description, dataset, and baseline,'' \emph{Proc. Workshop on
  Detection and Classification of Acoustic Scenes and Events {\rm (}DCASE{\rm
  )}}, pp. 69--73, 2018.

\bibitem{Salamon_PLoSOne2016_01}
J.~Salamon, J.~P. Bello, A.~Farnsworth, M.~Robbins, S.~Keen, H.~Klinck, and
  S.~Kelling, ``Towards the automatic classification of avian flight calls for
  bioacoustic monitoring,'' \emph{PLoS One}, vol.~11, no.~11, 2016.

\bibitem{Morfi_JASA2021_01}
V.~Morfi, R.~F. Lachlan, and D.~Stowell, ``Deep perceptual embeddings for
  unlabelled animal sound,'' \emph{The Journal of the Acoustical Society of
  America}, vol. 150, no.~2, pp. 2--11, 2021.

\bibitem{Morfi_DCASE2021_01}
V.~Morfi, I.~Nolasco, V.~Lostanlen, S.~Singh, A.~Strandburg-Peshkin, L.~Gill,
  H.~Pamu^^c5^^82a, D.~Benvent, and D.~Stowell, ``Few-shot bioacoustic event
  detection: A new task at the {DCASE} 2021 challenge,'' \emph{Proc. Workshop
  on Detection and Classification of Acoustic Scenes and Events {\rm
  (}DCASE{\rm )}}, pp. 145--149, 2021.

\bibitem{Valenti_IJCNN2017_01}
M.~Valenti, S.~Squartini, A.~Diment, G.~Parascandolo, and T.~Virtanen, ``A
  convolutional neural network approach for acoustic scene classification,''
  \emph{Proc. International Joint Conference on Neural Networks {\rm
  (}IJCNN{\rm )}}, pp. 1547--1554, 2017.

\bibitem{Liping_DCASE2018_01}
Y.~Liping, C.~Xinxing, and T.~Lianjie, ``Acoustic scene classification using
  multi-scale features,'' \emph{Proc. Workshop on Detection and Classification
  of Acoustic Scenes and Events {\rm (}DCASE{\rm )}}, pp. 29--33, 2018.

\bibitem{Tanabe_DCASE2018_01}
R.~Tanabe, T.~Endo, Y.~Nikaido, T.~Ichige, P.~Nguyen, Y.~Kawaguchi, and
  K.~Hamada, ``Multichannel acoustic scene classification by blind
  dereverberation, blind source separation, data augmentation, and model
  ensembling,'' \emph{Tech. Rep. DCASE Challenge 2018 Task5}, pp. 1--4, 2018.

\bibitem{Raveh_DCASE2018_01}
A.~Raveh and A.~Amar, ``Multi-channel audio classification with neural network
  using scattering transform,'' \emph{Tech. Rep. DCASE Challenge 2018 Task5},
  pp. 1--4, 2018.

\bibitem{Hershey_ICASSP2017_01}
S.~Hershey, S.~Chaudhuri, D.~P.~W. Ellis, J.~F. Gemmeke, A.~Jansen, R.~C.
  Moore, M.~Plakal, D.~Platt, R.~A. Saurous, B.~Seybold, M.~Slaney, R.~J.
  Weiss, and K.~Wilson, ``{CNN} architectures for large-scale audio
  classification,'' \emph{Proc. {IEEE} International Conference on Acoustics,
  Speech and Signal Processing {\rm (}ICASSP{\rm )}}, pp. 131--135, 2017.

\bibitem{Cakir_TASLP2017_01}
E.~\c{C}ak\i r, G.~Parascandolo, T.~Heittola, H.~Huttunen, and T.~Virtanen,
  ``Convolutional recurrent neural networks for polyphonic sound event
  detection,'' \emph{{IEEE/ACM} Transactions on Audio Speech and Language
  Processing}, vol.~25, no.~6, pp. 1291--1303, 2017.

\bibitem{Kong_TASLP2020_01}
Q.~Kong, Y.~Xu, W.~Wang, and M.~D. Plumbley, ``Sound event detection of weakly
  labelled data with {CNN}-{T}ransformer and automatic threshold
  optimization,'' \emph{{IEEE/ACM} Transactions on Audio Speech and Language
  Processing}, vol.~28, pp. 2450--2460, 2020.

\bibitem{Miyazaki_DCASE2020_01}
K.~Miyazaki, T.~Komatsu, T.~Hayashi, S.~Watanabe, T.~Toda, and K.~Takeda,
  ``Convolution-augmented transformer for semi-supervised sound event
  detection,'' \emph{Tech. Rep. DCASE Challenge 2020 Task4}, pp. 1--4, 2020.

\bibitem{Mesaros_EUSIPCO2011_01}
A.~Mesaros, T.~Heittola, and A.~Klapuri, ``Latent semantic analysis in sound
  event detection,'' \emph{Proc. European Signal Processing Conference {\rm
  (}EUSIPCO{\rm )}}, pp. 1307--1311, 2011.

\bibitem{Heittola_JASM2013_01}
T.~Heittola, A.~Mesaros, A.~Eronen, and T.~Virtanen, ``Context-dependent sound
  event detection,'' \emph{EURASIP Journal on Audio, Speech, and Music
  Processing}, vol. 2013, no.~1, 2013.

\bibitem{Imoto_IEICE2016_01}
K.~Imoto and S.~Shimauchi, ``Acoustic scene analysis based on hierarchical
  generative model of acoustic event sequence,'' \emph{IEICE Transactions on
  Information and Systems}, vol. E99-D, no.~10, pp. 2539--2549, 2016.

\bibitem{Imoto_TASLP2019_01}
K.~Imoto and N.~{Ono}, ``Acoustic topic model for scene analysis with
  intermittently missing observations,'' \emph{{IEEE/ACM} Transactions on Audio
  Speech and Language Processing}, vol.~27, no.~2, pp. 367--382, 2019.

\bibitem{Bear_INTERSPEECH2019_01}
H.~L. Bear, I.~Nolasco, and E.~Benetos, ``Towards joint sound scene and
  polyphonic sound event recognition,'' \emph{INTERSPEECH}, pp. 4594--4598,
  2019.

\bibitem{Tonami_IEICE2021_01}
N.~Tonami, K.~Imoto, R.~Yamanishi, and Y.~Yamashita, ``Joint analysis of sound
  events and acoustic scenes using multitask learning,'' \emph{IEICE
  Transactions on Information and Systems}, vol. E104-D, no.~02, pp. 294--301,
  2021.

\bibitem{Imoto_ICASSP2020_01}
K.~Imoto, N.~Tonami, Y.~Koizumi, M.~Yasuda, R.~Yamanishi, and Y.~Yamashita,
  ``Sound event detection by multitask learning of sound events and scenes with
  soft scene labels,'' \emph{Proc. {IEEE} International Conference on
  Acoustics, Speech and Signal Processing {\rm (}ICASSP{\rm )}}, pp. 621--625,
  2020.

\bibitem{Nada_APSIPA2021_01}
K.~Nada, K.~Imoto, R.~Iwamae, and T.~Tsuchiya, ``Multitask learning of acoustic
  scenes and events using dynamic weight adaptation based on multi--focal
  loss,'' \emph{Proc. {Asia-Pacific} Signal and Information Processing
  Association Annual Summit and Conference {\rm (}APSIPA ASC{\rm )}}, pp.
  1156--1160, 2021.

\bibitem{Liu_CVPR2019_01}
S.~Liu, E.~Johns, and A.~J. Davison, ``End-to-end multi-task learning with
  attention,'' \emph{{IEEE/CVF} Conference on Computer Vision and Pattern
  Recognition {\rm (}CVPR{\rm )}}, pp. 1871--1880, 2019.

\bibitem{Lin_ICCV2017_01}
T.~Y. Lin, P.~Goyal, R.~Girshick, K.~He, and P.~Doll\'{a}r, ``Focal loss for
  dense object detection,'' \emph{Proc. {IEEE} International Conference on
  Computer Vision {\rm (}ICCV{\rm )}}, pp. 2980--2988, 2017.

\bibitem{Noh_Sensors2020_01}
K.~Noh and J.~H. Chang, ``Joint optimization of deep neural network-based
  dereverberation and beamforming for sound event detection in multi-channel
  environments,'' \emph{Sensors}, vol.~20, no.~7, pp. 1--13, 2020.

\bibitem{Mesaros_EUSIPCO2016_01}
A.~Mesaros, T.~Heittola, and T.~Virtanen, ``{TUT} database for acoustic scene
  classification and sound event detection,'' \emph{Proc. European Signal
  Processing Conference {\rm (}EUSIPCO{\rm )}}, pp. 1128--1132, 2016.

\bibitem{Mesaros_DCASE2017_01}
A.~Mesaros, T.~Heittola, A.~Diment, B.~Elizalde, A.~Shah, B.~Raj, and
  T.~Virtanen, ``{DCASE} 2017 challenge setup: Tasks, datasets and baseline
  system,'' \emph{Proc. Workshop on Detection and Classification of Acoustic
  Scenes and Events {\rm (}DCASE{\rm )}}, pp. 85--92, 2017.

\bibitem{Imoto_dataset2019_01}
\url{https://www.ksuke.net/dataset}.

\bibitem{Imoto_ICASSP2021_01}
K.~Imoto, S.~Mishima, Y.~Arai, and R.~Kondo, ``Impact of sound duration and
  inactive frames on sound event detection performance,'' \emph{Proc. {IEEE}
  International Conference on Acoustics, Speech and Signal Processing {\rm
  (}ICASSP{\rm )}}, pp. 875--879, 2021.

\bibitem{Liu_ICLR2020_01}
L.~Liu, H.~Jiang, P.~He, W.~Chen, X.~Liu, J.~Gao, and J.~Han, ``On the variance
  of the adaptive learning rate and beyond,'' \emph{Proc. International
  Conference on Learning Representations {\rm (}ICLR{\rm )}}, pp. 1--13, 2020.

\end{thebibliography}
\end{document}